\documentclass[iop]{emulateapj}
\usepackage{apjfonts}
\journalinfo{The Astrophysical Journal, 2014, in press}

\def\RSUN{$R_{\sun}$}

\def\kms{$\rm km~s^{-1}$}

\def\C3{$\rm C~III$ }

\def\Si12{$\rm Si~XII$ }

\shorttitle{Comet Lovejoy as a Probe of the Solar Corona}
\shortauthors{Raymond et al.}

\begin{document}

\title{The Solar Corona as probed by Comet Lovejoy (C/2011 W3)}

\author{J. C. Raymond,  P. I. McCauley, S. R. Cranmer}
\affil{Harvard-Smithsonian Center for Astrophysics, 60 Garden Street,
Cambridge, MA  02138}
\and
\author{C. Downs}
\affil{Predictive Sciences, Inc., 990 Mesa Rim Road,
Suite 170, San Diego, CA, 92121}

\begin{abstract}
EUV images of Comet Lovejoy (C/2011 W3) from the AIA show striations related to the magnetic
field structure in both open and closed magnetic regions.  The brightness contrast 
implies coronal density contrasts of at least a factor of 6 between neighboring flux 
tubes over scales of a few thousand km.  These density structures
imply variations in the Alfv\'{e}n speed on a similar scale.  They will drastically
affect the propagation and dissipation of Alfv\'{e}n waves, and that should be taken
into account in models of coronal heating and solar wind acceleration.  In each
striation, the cometary emission moves along the magnetic field and broadens
with time.  The speed and the rate of broadening are related to the parallel
and perpendicular components of the velocities of the cometary neutrals when they
become ionized.  We use an MHD model of the coronal magnetic field and the
theory of pickup ions to compare the measurements with theoretical predictions,
in particular with the energy lost to Alfv\'{e}n waves as the cometary ions isotropize.
\end{abstract}


\section{Introduction}

EUV Images of Comet Lovejoy (C/2011 W3) obtained with the 
Atmospheric Imaging Assembly (AIA) instrument
on the Solar Dynamics Observatory (SDO) show a tail that consists of
a remarkable series of striations a few arcseconds across
at an angle to the comet's path.  Those striations change direction abruptly 
along the trajectory, and  \cite{downs13} showed that they
align with the magnetic field based on magnetohydrodynamic (MHD) simulations with
the thermodynamic version of the Magnetohydrodynamic Algorithm on a
Sphere (MAS)  code.  \cite{mccauley13} showed that
the positions, lengths and relative intensities of the tail
as seen in the various AIA filters agree with the theory
that the emission arises from cometary
gas as it is ionized through successive ionization states
\citep{bryanspesnell12}, with most of the wavelength bands
dominated by emission from oxygen.  \cite{mccauley13}  were able to 
determine the outgassing rate of the comet and the relative
abundances of carbon, oxygen and iron.

In this paper we investigate the striations further.  We use them to
probe the structure of the corona at a single position along the comet's
trajectory, rather than relying on a line-of-sight average as
is typical of most remote sensing observations. In
particular, we interpret the striations as magnetic flux tubes containing
relatively high densities such that neutral H and O atoms from
the comet travel with the comet through the low density plasma
between the dense tubes.  When they reach a denser
region, they become ionized and confined to the flux tube, traveling
along it but unable to cross field lines.  The lack of emission between
the striations requires that few of the neutrals released from the comet
in those regions become ionized in the time it takes the
comet to cross, while most of the neutrals
released within a striation become ionized.  That implies ionization
times, and hence densities, for the high and low density flux tubes. 

Fine structure in the corona has been seen earlier in processed 
eclipse images as filaments 1000--5000 km across
\citep{novemberkoutchmy96, woo07}.  While the observed contrast level
is small, the corona is optically thin and the path length is large,
so that November \& Koutchmy estimated a  
local density contrast of $\Delta n/n = 100$\%.  We show that the density 
contrast we derive is compatible with the eclipse observations.  On 
larger scales, polar plumes stand out above the background \citep{tian11},
and they have similar density contrasts \citep{young99}.

The small scale density structure has important implications for
the propagation and dissipation of Alfv\'{e}n waves.  Sharp gradients
in Alfv\'{e}n speed cause rapid dispersion and phase mixing, converting
the waves to shorter wavelength modes that can damp rapidly \citep{leeroberts86,
evans12}.  Density gradients can also give rise to entirely different 
modes such as drift waves, especially if velocity shear is present, and
they might play a role in coronal heating \citep{saleem12}.  If 
the solar wind mass flux remains roughly constant across the
striations, then regions of high (low) density will be correlated
with regions of low (high) radial velocity.  In such highly sheared
regions, it has been shown that some fraction of an incoming Alfv\'{e}n
wave train can be transformed into a compressive, longitudinal form similar
to a fast-mode MHD wave (e.g., \cite{nakariakov98, gogoberidze07, hollweg13})

We compare
the observed scale of the striations with predictions for the 
energy-containing scales (the scales of the waves that contain most of the
power and cascade to smaller scales to deposit their energy), 
which are central to models of coronal heating \citep{abramenko, 
cranmervB05, hollweg10}.  While we concentrate on an open field region of
slow solar wind, we also consider a closed field region.

An important aspect of the interaction of the oxygen ions with
the magnetic field is that they behave as pickup ions \citep{moebius}.  We
use the velocity of the comet and the direction of the magnetic field
from the MHD model of \cite{downs13} to calculate the velocity components
parallel and perpendicular to the magnetic field.  To compute the speed of
the cometary plasma along the field direction and its rate of
spreading, we assume that the particles relax to a bispherical
distribution according to the theory of \cite{williamszank}.  We then
compare those predictions with the observations.

In the following sections, we briefly summarize the properties of Comet Lovejoy,
discuss the AIA observations and analyze their implications for the coronal
density structure, calculate the predicted velocity and velocity spread along
the magnetic field, discuss the implications for MHD wave propagation and summarize
our findings.

\section{Comet Lovejoy}  

Comet Lovejoy is the largest member of the Kreutz family of sungrazing comets 
\citep{marsden05} seen in several decades.  It reached perihelion
at about 1.2 \RSUN on 16 Dec.~2011 and survived for 1.6 days afterwards before the
nucleus disintegrated.  The disruption was probably due to thermal stresses in the
comet interior \citep{sekaninachodas12}.  The diameter of the nucleus was estimated
to be around 600 m before it approached the Sun \citep{mccauley13}.

At heliocentric distances above about 3 \RSUN, Comet Lovejoy displayed a bright dust tail,
but at smaller distances the dust grains sublimate quite rapidly and the dust tail
disappeared \citep{sekaninachodas12}.  However, a different tail visible in the UV and
EUV channels of the AIA instrument was seen close to the Sun.  As described by 
\cite{bryanspesnell12}, most of the AIA wavelength bands contain some lines of ionized
oxygen along with the lines of highly ionized iron they were designed to capture.  
As the oxygen produced by photodissociation of water from the comet proceeds
through the successive ionization stages, it emits photons in those bands.  
\cite{mccauley13} showed that at coronal temperatures, each oxygen ion emits a fixed number of
photons in each AIA band before it is ionized, so that it is simple to compute the number
of O atoms released per second from the brightness in the AIA images.  The AIA 1600 \AA\/ band is
dominated by C~IV rather than any O ion, and some emission from Fe in the 
AIA 171 \AA\/ band was indicated by comparison with the 131 \AA\/ and 193 \AA\/ bands,
which are dominated by emission from the same O ions present in the 171 \AA\/ band 
(O V and O VI).
It was possible to match the relative count rates in the different bands to within about
a factor of 2, derive the relative abundances of C, O and Fe, and determine the outgassing
rate.  At the time of peak brightness, at about 00:46 UT, \cite{mccauley13} derived
an outgassing rate of about $3 \times 10^{32}$ oxygen atoms per second, for a mass loss rate
of about $9 \times 10^6$ g/s.

\section{Observations}
\label{obs}

The AIA observations are presented in \cite{mccauley13} and \cite{downs13}.
A set of images was obtained every 12 seconds with 0.6$^\prime$$^\prime$ pixels
for a spatial resolution of
1.5$^\prime$$^\prime$ \citep{lemen12}.  For the analysis here, we confine ourselves to the 171 \AA\/ band.
The striations can be seen most clearly in that band because of its large effective
area and the strong O V, O VI and Fe IX lines in the bandpass.  In addition,
the ionization times of those ions are longer than the ionization times
of the lower ions that dominate the other wavelength bands, so the striations
are longer in the 171 \AA\/ band.  We analyze images obtained near 00:46 UT, when the
comet was about 1.3 \RSUN from Sun center.

\begin{figure}
\epsscale{1.13}
\plotone{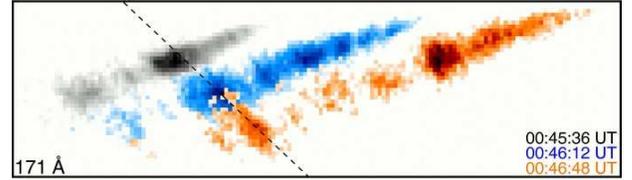}
\caption{Three co-aligned AIA images in the 171\AA\/ band taken 36 seconds apart.  They
show the centroid motion and the spreading of the emitting plasma along the
magnetic field.  The background coronal emission has been removed based on
pre- and post-comet images. (Adapted from \cite{mccauley13})
\label{mccauley}}
\end{figure}
Figure~\ref{mccauley}, which reproduces part of Figure 3 of \cite{mccauley13}, is a superposition
of three images in the AIA 171 \AA\/ band at intervals of 36 seconds.  It clearly shows how
the emission in each individual striation moves along the magnetic field and spreads
along the field with time.  In order to estimate the densities inside and outside
of the striations, we will use the intensity contrast.  In order to study the
pickup ion behavior of the oxygen ions, we will use the speed along the magnetic
field and the rate of spreading.  

\begin{figure}
\epsscale{1.17}
\plotone{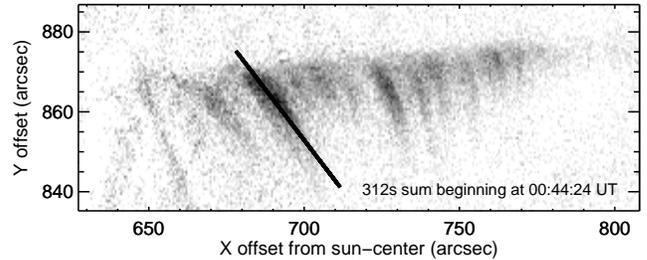}
\caption{The sum of 26 2-second AIA 171 \AA\/ images taken over 312 seconds.
The black line indicates the direction adopted for the striation in the 
subsequent figures.  Intensities are extracted parallel
and perpendicular to this direction.  Note that the direction changes
with position along the X-axis and that some of the striations curve, so 
that the contrast between on- and off-striation regions is washed out at
X offsets above about 730$^\prime$$^\prime$. 
\label{images}}
\end{figure}
Figure~\ref{images} shows the sum of 26 2-sec exposures obtained over 312 seconds, beginning
at 00:44:24 UT during comet egress.  Since the ionized plasma from the comet moves
along magnetic field lines, the striations show the field line structure.  The curvature
seen in some of the striations could be partly due to intrinsic curvature of the field
lines, but it may be largely due to relaxation of the field lines after they are perturbed
by the ram pressure of the cometary gas.

\begin{figure}
\epsscale{1.17}
\plotone{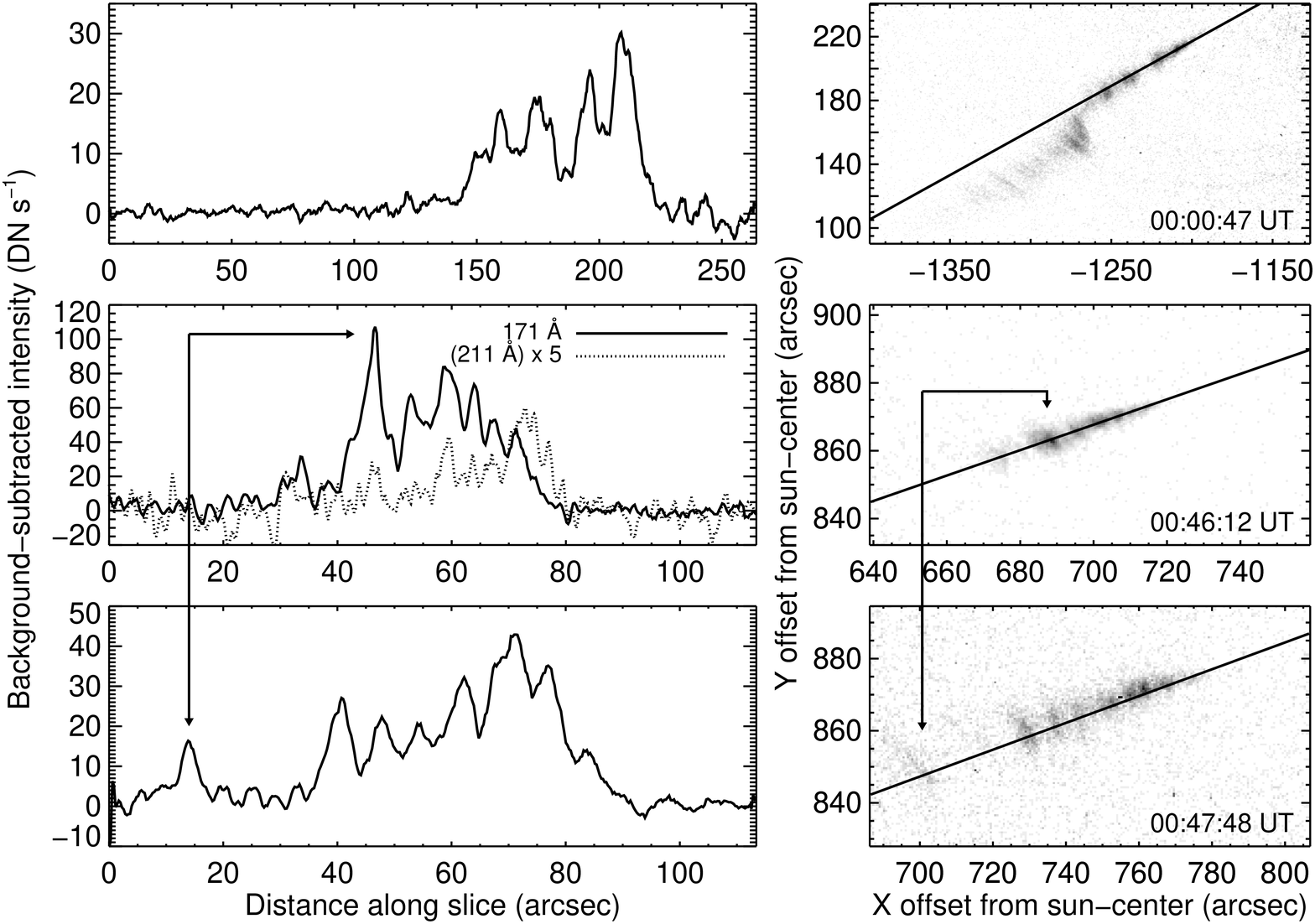}
\caption{Intensity cuts perpendicular to the striations during ingress (Top row)
and at two times during egress (Bottom two rows).  The
arrows in the bottom panels indicate the bright blob chosen for the striation direction in 
Figure~\ref{images} and the profiles shown in Figures~\ref{width} and \ref{profile}. 
\label{intensity}}
\end{figure}
Figure~\ref{intensity} shows images of the comet in the right panels and corresponding
plots of the intensities along the lines cutting across the striations in the left 
panels.  The top row shows
the intensities for an image during ingress while the lower two rows show the intensities
for two times during egress.
The intensity contrast is a factor of 2 to 3 and the separation between striations
is about 5$^\prime$$^\prime$, or 4000 km, in the egress images.  That contrast may be 
something of an underestimate due to the instrumental resolution of 1.5$^\prime$$^\prime$, 
so the factor of 2 to 3 is a lower limit.  The filling factor of the striations is about
1/2, but that may also be influenced by the instrumental resolution.  The amplitude of the
brightness fluctuations in the ingress image is similar, but the separations are larger and
less regular.  The analysis in the following sections will concentrate on the egress 
observations, but we will return briefly to the ingress observation below.

\begin{figure}
\epsscale{1.11}
\plotone{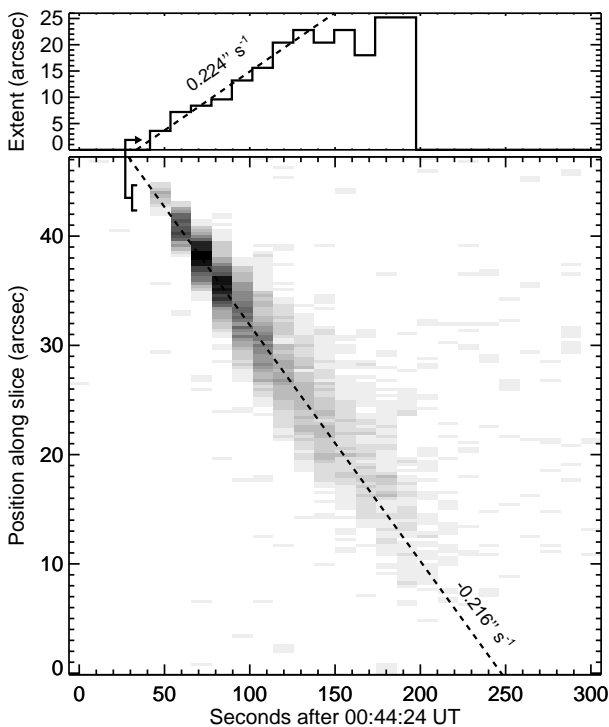}
\caption{The FWHM extent (top panel) and centroid (lower panel) of the emission
in the AIA 171 \AA\/ band as a function of time, measured along the striation
as shown in Figure~\ref{images}.  These characterize the rate at which the emission
region spreads (top) and the speed at which the plasma travels along the magnetic
field (bottom).
\label{width}}
\end{figure}
The top panel in Figure~\ref{width} shows the extent of the brightest striation 
(indicated by the arrows in Figure~\ref{intensity}) along its axis at each time.  
The values shown are FWHM widths of the
brightness distribution along the filament.  The width grows with time due to the spread of the
ion velocities along the magnetic field.  At later times the width saturates as the emission
becomes very faint. The slope of the width vs. time plot is 0.224$^\prime$$^\prime$/second, or 
159 km/s in the plane of the sky.  The lower panel shows the centroid position of the emission.
The slope gives the plane-of-sky projection of the velocity along the magnetic field.  The 
value of 0.216$^\prime$$^\prime$/sec corresponds to 154 km/s.  From the magnetic field
vector and the line-of-sight direction \citep{downs13}, we find that the measured 
(projected) values are 0.82 times
the actual deprojected values.  Therefore, $V_\parallel = 183$ \kms\/ and the
emitting region spreads at 200 \kms.

\begin{figure}
\epsscale{1.13}
\plotone{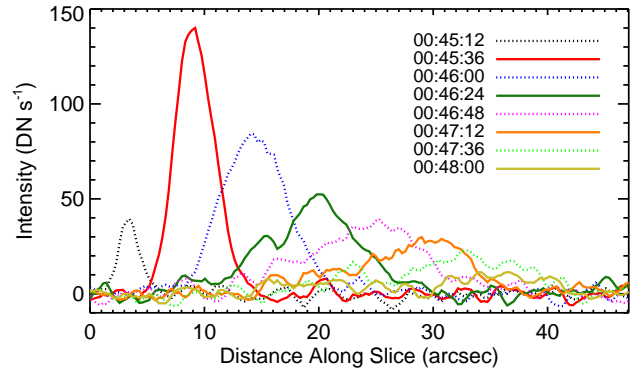}
\caption{Profiles of the AIA intensity along the striation indicated
in Figure~\ref{images}.  The motion along the striation and its increasing
extent are apparent.  The profile at 00:45:36 is close to Gaussian, while
subsequent profiles are more skewed.
\label{profile}}
\end{figure}

\section{Analysis}

We have chosen the observations during egress at about 00:44.5 UT for detailed analysis.  
According to the MAS models, this is an open field region where the coronal plasma 
flows at 43 \kms.  That speed indicates that this is a slow solar wind region.

\subsection{Density Contrast}

The striations are clearly aligned with the local magnetic field \citep{mccauley13,
downs13}.  They are quasi-periodic, which might suggest a periodic source from the comet,
and Figure 7 of \cite{mccauley13} shows that the outgassing rate varies with time.  However,
the comet travels from one striation to the next in about 9 seconds, and it is not
plausible that the comet rotates that rapidly.  Nine seconds is much less than the critical
rotation period for disruption of around half an hour \citep{jewitt97} and much less than observed comet
rotation periods of hours to days.  Instead, the striations arise
because neutrals are not tied to the magnetic field, while ions are.  The comet
produces water at a rate of $3 \times 10^{32}$ molecules per second near the brightness 
peak at 00:49 UT \citep{mccauley13}.  Photodissociation quickly splits the molecules
into H and O atoms which share the comet's motion and form a cloud that expands at 
a few \kms\, \citep{combi96} as it travels at 530 \kms through the corona.  The atoms travel 
freely until they become ionized and trapped on a field line.

We can obtain an upper limit to the density between striations from the intensity
contrast and the condition that few of the cometary atoms are ionized in the
inter-striation plasma.  The comet crosses one of these 4$^\prime$$^\prime$ regions
in about 5 seconds.  The ionization rate coefficients of H and O atoms are $2.9\times 10^{-8}$
and $8.7 \times 10^{-8}~n_e ~~\rm cm^3~s^{-1}$, respectively,
at temperatures of 1.5 to 2 MK according to the rates of \cite{dere07} as
given by version 7.1 of CHIANTI \citep{dere97, landi13}.  Ionization can also occur by way
of charge transfer with protons in the corona.  For H and O atoms the rate coefficients 
$n_p V_{\rm comet} \sigma$ are $8.5\times 10^{-8}$ and $3.5 \times 10^{-8} ~n_p~~\rm cm^3~s^{-1}$ 
for a speed of 530 \kms based on cross sections
of \cite{schultz08} and \cite{phaneuf87}.  Thus the total ionization rate coefficients for 
both H and O are about $1.2 \times 10^{-7}~\rm cm^3~s^{-1}$. The condition that fewer 
than 30\% of the atoms are ionized between striations ($t_{\rm cross} < 0.3 t_{\rm ion}$)  
translates into an upper limit of $2.7 \times 10^6~\rm cm^{-3}$ on the coronal density
between the striations.

On the other hand, most of the atoms are ionized in the striations.
In order for 90\% of the atoms from the comet to become ionized within a striation,
the crossing time has to be at least twice the ionization time,
$t_{\rm cross} > 2 t_{\rm ion}$.
That implies a density of at least $1.7 \times 10^7~\rm cm^{-3}$ within the striations.  Thus the density 
contrast must be at least a factor of 6.  We note that if the striations occupy about
half the space surrounding the comet and have densities near this limit, the average density is around 
$8 \times 10^6~\rm cm^{-3}$.  The MAS model predicts a density of 
$2.6 \times 10^6~\rm cm^{-3}$ at the point along the comet's path corresponding to the 
AIA image at 00:45:36 UT, but
we will argue below that the the magnetic field direction in is better agreement with
the position which the comet reached 2--3 minutes earlier, and the predicted density there is
$6.7 \times 10^6~\rm cm^{-3}$, in reasonable agreement with our estimated average density.
This shift of 2 or 3 minutes corresponds to an offset of about 4 degrees in solar longitude,
which is quite reasonable given the uncertainty in the magnetic field at the solar surface
used by the MAS model.

Another estimate of the density in the striations can be obtained from the emission
measure ($EM$), the filling factor ($f$), and an estimate of the length scale through the 
corona ($L$), $n_e = (EM/(Lf))^{1/2}$.  \cite{mccauley13} obtained an emission
measure from the AIA images just outside the comet's tail of $1.9 \times 10^{26}~\rm cm^{-5}$.
The filling factor appears to be about 0.5 based on the fraction of the tail
included in the striations, and the length scale
is roughly 0.6 \RSUN.  Thus we estimate $n_e = 1.3 \times 10^8~\rm cm^{-3}$.  This is clearly an average
for the high density regions along the line of sight.  Since the comet was not too far
from the plane of the sky at that time, it may be a
reasonable estimate, but the observations pertain to an open field region, and closed
field loops along the line of sight could easily dominate the emission measure.  Therefore,
we take this to be an upper limit.  This limit agrees well with the estimate of 
$1.4 \times 10^8~\rm cm^{-3}$ derived from the ionization time from O III to O VI by 
\cite{mccauley13}.  The density derived from the ionization time includes the electrons 
liberated by ionizing hydrogen and oxygen from the comet, so we conclude that 
$1.4 \times 10^8~\rm cm^{-3}$ is a solid upper limit to the density in the striations 
before the comet's arrival, while a density of $1.7 \times 10^7~\rm cm^{-3}$ from
the density contrast is a lower limit.

The real situation is complex, of course.  The density of cometary atoms is large
enough that, once they begin to ionize, they significantly increase the total density.
However, the ions and electrons move away from the cloud of neutrals at over 150 \kms.
Another complication is the uncertainty in the electron temperature.  Ionization
by electrons continues as long as the electrons remain hot.  Each ionization requires
the ionization potential of 13.6 eV, and roughly 16 eV is lost in photons produced
by collisional excitation.  Moreover, the total thermal energy is shared among
an increased number of electrons, rapidly reducing the temperature, so that each coronal
electron can ionize only 3 or 4 cometary atoms unless additional heat is supplied.
There is plenty of energy available in the cometary ions, so it is likely that the electron temperature 
is maintained by Coulomb collisions with ions.  A detailed calculation is beyond the scope of this paper,
but an electron temperature of 1--2 MK is consistent with the ionization times inferred from the
relative intensities in the different AIA bands and with the X-rays detected with the XRT instrument
on {\it Hinode} \citep{mccauley13}.

\subsection{Nature of the Density Structure}

The cause of the inhomogeneous coronal density structure is not yet clear.
One possibility is that the variations in density originate at the
footpoints of magnetic flux tubes and propagate upward.
Intensity oscillations measured with off-limb EUV imaging and
spectroscopy imply the presence of compressible MHD waves that appear
channeled along polar plumes (e.g., \cite{deforestgurman, ofman99, krishnaprasad}).
These fluctuations have periods of order 10--20 minutes, and if they
were slow-mode MHD waves ($V_{\rm ph} \approx v_{\rm wind} +
c_{s} \approx 200$ km s$^{-1}$), they could have wavelengths of
order 0.2--0.4 $R_{\odot}$.
These temporal and spatial scales are larger than the resolved
passage through the striations, so the comet would be expected to
``sample'' such oscillations as quasi-static structures.
The measured off-limb density fluctuation ratio $\delta \rho / \rho_{0}$
is only of order 0.03--0.1, but line-of-sight integration effects
are likely to wash out the true local fluctuation amplitude.

Another type of proposed upflowing density variation is the jet-
or piston-like motion associated with Type II spicules.
These features have been suggested as a major source of mass for the
corona and solar wind \citep{depontieu11,pereira12},
though \cite{klimchuk12} argues that these upflows are probably a
minor contributor.
A related possibility is that mass is injected into specific flux
tubes in the slow solar wind by reconnection of open and closed
magnetic fields in the low corona \citep{schwadron99, antiochos11}.
That model was advanced specifically for the slow solar
wind, but we see similar striations in fast wind regions, such as
just before the kink in the upper panel of Figure 3.  Reconnection
between open and closed magnetic flux also occurs in coronal holes,
but one might expect different scales lengths and filling factors.
Neither of the above impulsive-driving scenarios makes a clear
prediction as to the size or filling factor of the high and low
density regions.

\begin{figure}
\epsscale{1.13}
\plotone{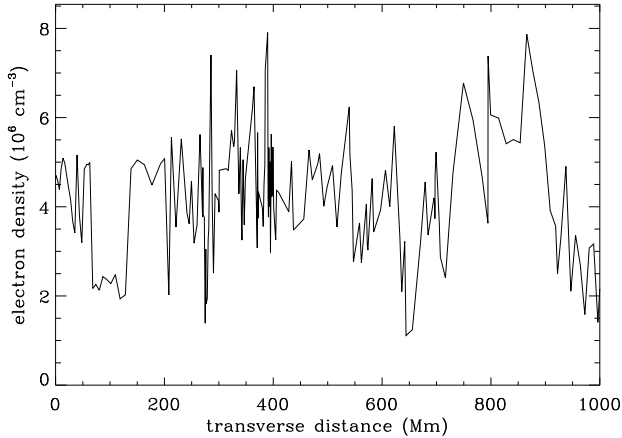}
\caption{Modeled electron density plotted as a function of transverse
distance along a set of magnetic flux tubes rooted in a low-latitude
quiet Sun region observed in 2003 September by SOLIS.  The
densities were sampled at a constant radius of $r=1.3 \, R_{\odot}$
\citep[see][]{cranmer13}.
\label{densmodel}}
\end{figure}

Still another picture is that variations in the expansion factor of
different flux tubes give rise to different rates of Alfv\'{e}n wave
reflection, turbulent cascade, and coronal heating.  This in turn
causes differences in the flow speeds and densities in neighboring
flux tubes. \cite{cranmer13} found significant inter-tube variations
in a model containing a dense grid of flux tubes that connected
a high-resolution quiet-Sun magnetogram with the ecliptic plane.
The model solved time-steady equations of mass, momentum, and
energy conservation along each flux tube using a self-consistent
description of non-WKB Alfv\'{e}n wave reflection and turbulent
heating.
Each of the individual flux tube models was given identical lower
boundary conditions at the photosphere, and they differed only in
the radial dependence of magnetic field strength.
Figure~\ref{densmodel} shows a cut through the modeled set of flux tubes at a
constant radius of $r = 1.3 \, R_{\odot}$.
At that height, the mean separation between neighboring flux
tubes was 5100 km, and roughly 45\% of the flux tubes exhibited
nearest-neighbor separations less than the observed striation
size of 4000 km.
The minimum, mean, and maximum electron densities in this set of
models were $1.1 \times 10^{6}$, $4.2 \times 10^{6}$, and
$7.9 \times 10^{6}$ cm$^{-3}$.
The ratio of the standard deviation to the mean density was 0.32,
which implies a representative contrast factor of order 3.
The distribution of transverse scale sizes is roughly a power law,
with no dominant scale that separates the densest flux tubes from
their more rarefied surroundings.

It is also within the realm of possibility that a sort of
filamentation instability operates, based perhaps on the focusing of
Alfv\'{e}n waves into the lower Alfv\'{e}n speed regions corresponding
to higher density flux tubes.
In quasi-steady flow models such as those of \cite{champeaux},
enhanced heating below the critical point in those flux tubes would
further increase the density in those flux tubes.
Nevertheless, it does appear to be possible that the observed
striations could be the result (at least in part) of natural
variations in the time-steady coronal heating and wind
acceleration.
In other words, it is not surprising that a highly structured
magnetic field gives rise to a highly structured coronal plasma.

\subsection{Pickup Ion Behavior}

The cloud of neutral hydrogen and oxygen atoms produced by
photodissociation of water moves with the comet and expands
at a few \kms.  When the atoms become ionized as they travel
through the coronal magnetic field they behave as pickup
ions \citep{williamszank, isenberglee}.  Initially they have velocity components
$V_{\|} = V_{\rm comet} \cos \theta$ and $V_{\bot} = V_{\rm comet} \sin \theta$,
where $\theta$ is the angle between the comet's trajectory
and the magnetic field.

Since $V_{\bot}$ is the same for all the newly formed ions,
they form a ring distribution in velocity space.  The ring
distribution is unstable, and it evolves into a bispherical
shell in velocity space by emitting and absorbing Alfv\'{e}n
waves on a time scale of a few gyroperiods \citep{williamszank}.  The
bispherical shell is described in detail by Williams \&
Zank, and in particular the energy of the bispherical
distribution is given by their Equation 4 as
\begin{equation}
E_{BD\pm} = \frac{nm\pi v_\pm^2}{a_T} \left[ \frac{V_{\|}}{v_\pm}
(V_{\|} V_A \pm V_A^2 \mp v_\pm^2) + (v_\pm^2 - V_A)^2 \right]
\end{equation}
where $v_\pm^2 \equiv ~ V_{\bot}^2 ~+~ (V_{\|}~\pm~V_A)^2$,
the areas of the sections of the bispherical shell are
$a_\pm ~\equiv ~  2 \pi v_\pm (v_\pm ~\mp~ V_{\|} ~-~ V_A)$,
$a_T ~\equiv~ a_+ ~+~ a_-$, and of course $V_A$ is the Alfv\'{e}n speed.
Under the assumption that the ions uniformly cover the bispherical
shell, Equation 3 of \cite{williamszank} gives the bulk speed 
along the magnetic field.

To compare the predicted pickup ion behavior with observations,
we use the MAS model predictions for the magnetic field direction
and the Alfv\'{e}n speed.  That provides the parameters for the
bispherical distribution and for the projection of the velocities
onto the plane of the sky.  We consider the position corresponding to the
observation at 00:44.5 UT shown in Figures 1 through 5, where the
comet was moving at 527 \kms, and where the high outgassing rate
provided us with the best data.

The MAS model magnetic field at the 00:44.5 UT
position lies only 3$^\circ$ from the direction perpendicular to the comet's motion,
so that the predicted speed along the striation is only 24 \kms.  However,
the predicted direction of the field is changing rapidly in that
region.  In addition, the position of that rapid change is somewhat uncertain
as a result of the uncertainty in the solar surface magnetic field that
serves as a boundary condition, especially because the relevant
region lies just behind the solar limb, and no magnetic
field measurements exist for the back side of the Sun.  Therefore,
we also consider the coronal field predicted for a region that the
comet passed 2.5 minutes earlier, at 00:41.9 UT.  At that position the angle
between the comet's orbit and the field was larger so that the predicted
$V_{\|}$, while still lower than the observed value, is much
more reasonable.

Table 1 lists the velocity components of the comet's motion and
the components of the magnetic field in the Carrington coordinate
system, along with the density and Aflv\'{e}n speed, as predicted
by the MAS model.
It also compares the observed and predicted values of
the bulk velocity and expansion velocity along the field in
the plane of the sky ($V_{s,POS}$ and $V_{exp,POS}$, respectively)
and the total kinetic energy (after deprojecting
the plane-of-the-sky velocities) with the predicted energy of the
bispherical distribution ($E_0$ and $E_{BD}$, respectively).
\begin{table}
\begin{center}

\vspace*{0.07in}
\centerline{\bf Table 1}

\centerline{Model and Observed Parameters}

\vspace*{0.07in}
\begin{tabular}{l r r r }
\hline
\hline
                     & Obs  &    Model & Model \\
                     &      &    00:45.4 UT    & 00:41.9 UT  \\
\hline
$V_{com,x}$          &      &  392  &   385  \\
$V_{com,y}$          &      &  347  &   362  \\
$V_{com,z}$          &      &   31  &    45  \\
$B_x$                &      &   0.29&   0.34 \\
$B_y$                &      &  -0.23&  -0.11 \\
$B_z$                &      &  -0.56&  -0.50 \\
$n_e$                &      & $2.6 \times 10^6$& $6.7 \times 10^6$ \\
$V_A$                &      &    878&  480  \\
$V_{s,POS}$          & 154   &    24 &   93  \\
$V_{exp,POS}$        & 159   &   120 &  188  \\
$E_0$                &      &  21880& 22500 \\
$E_{BD}$             & 8780 &  11840& 13600 \\
\hline
\end{tabular}
\vspace*{0.07in}
\centerline{%
velocities in \kms, magnetic fields in G,}

\vspace*{-0.07in}
\centerline{density in $\rm cm^{-3}$, and energies in eV per O atom.}
\end{center}
\end{table}

The predicted motion along the field for the comet position at 00:44.5 UT
is far too small to match the observation, while that at 00:41.9 UT is closer
to the observed value.  This corresponds to a shift of about 4 degrees in 
solar longitude. It still does not agree very well, and we consider
that to be an indication that the angle between the comet trajectory and
the field is larger than that given by the model.  An increase of about 4
degrees is required, depending on the projection onto the plane of the
sky, and that is not unreasonable given the expected accuracy of the model.

The predicted expansion speed along the field line for the model
parameters at 00:41.9 UT is somewhat larger than observed.  We consider
the comparison of the energy in the bispherical distribution to the
initial kinetic energy to be the strongest test of the pickup ion
theory.  The observed value of $E_{BD}$ depends on the deprojection
of the centroid and expansion speeds, but those are not large corrections.
The theory outlined above, taken with the Alfv\'{e}n speed and angle
between the comet motion and magnetic field direction, predicts
$E_{BD} / E_0$ = 0.60, while the observed value is 0.39.

Coulomb collisions will transfer some energy from the oxygen ions to
protons and electrons on a time scale comparable to the duration of
our measurements.  Table 2 lists the ionization times ($t_{\rm ion}$), 
gyration periods ($t_{\rm Larmor}$) and Coulomb collision times for 
transfer of energy among oxygen ions ($t_{\rm O-O}$) and between oxygen 
ions and protons ($t_{\rm O-H}$) for the relevant O ions.  Ionization times
are from the work of \cite{dere07}, while Coulomb collision times
are from \cite{spitzer}.  Coulomb collision times for O ions and electrons
are comparable to those for O ions and protons, but we do not list them
because we do not yet have firm predictions for the electron temperature.
We assume a magnetic field of 0.6 G based on the MAS model and a density
of $10^8$ cm$^{-3}$ based on the relative positions and lengths of
the striations in the different AIA filters (McCauley et al.\  2013).
It is apparent that Coulomb losses might account for the the difference
between the observed and predicted values of $E_{BD}$  and $E_0$, but a
more detailed calculation of the evolution of the cloud of cometary
ions is required.
\begin{table}
\begin{center}

\vspace*{0.07in}
\centerline{\bf Table 2}

\centerline{Collision and Gyration Time Scales}

\vspace*{0.07in}
\begin{tabular}{l r r r r }
\hline
\hline
         & $t_{\rm ion}$  & $t_{\rm Larmor}$ & $t_{\rm O-O}$ & $t_{\rm O-H}$ \\
\hline
$\rm O^+$     & 0.37  &  0.018  & 55600  & 1740  \\
$\rm O^{2+}$  & 0.89  &  0.009  & 3480   &  435  \\
$\rm O^{3+}$  & 2.6   &  0.006  &  686   &  193  \\
$\rm O^{4+}$  & 9.6   &  0.0045 &  217   &  109  \\
$\rm O^{5+}$  & 33.   &  0.0036 &   89   &   69  \\
\hline

\end{tabular}
\vspace*{0.07in}
\centerline{Times in seconds}
\end{center}
\end{table}

\section{Discussion}

We find that neighboring flux tubes show density variations of at least a factor of 6
on a scale of few thousand km in an open field region at a height of about 1.3 \RSUN.
While we have done a detailed analysis at only one position, similar striations are seen all along
along the comet's path.  From Figure 1 of \cite{mccauley13}, there is an indication
of wider separations at larger heliocentric heights.  Earlier in the egress, when the comet
was at a lower height, the tail was narrower and more continuous, presumably as a
result of higher density in that region.  The structure resembles that
inferred from an edge-enhanced white light eclipse image by \cite{novemberkoutchmy96},
who also inferred a substantial density contrast. 

It is unclear to what extent the small-scale striations elucidated
in this paper relate to the larger-scale {\em polar plumes} that
have been long known to exist in coronal holes.
Plumes are bright ray-like features that trace out open magnetic
field lines and exhibit a strong intensity contrast in off-limb images
(e.g., \cite{newkirkharvey, ahmadwithbroe, suess82}).
Observational attempts to measure the density contrast between plumes
and the more tenuous interplume corona have yielded different answers
depending on the exact diagnostic techniques used.
At heliocentric radii around 1.1--1.3 $R_{\odot}$, several analyses
of the line-of-sight EUV and white-light emission have given density
contrast ratios of order 3--6 (e.g., Saito 1965; Young et al.\  1999).
However, at similar heights, Orrall et al.\  (1990) combined the EUV
and white-light data to estimate a statistical irregularity ratio
$\langle n^{2} \rangle / \langle n \rangle^{2}$.
They inferred the presence of substantially higher contrast ratios
of order 20--60.
Orrall et al.\  concluded that there must be density fluctuations
on spatial scales below their resolution of 5\arcsec--60\arcsec.

The inferred density contrast between flux tubes indicates variations
in the Alfv\'{e}n speed of at least factors of 2--3.
This points to the likelihood that large-scale MHD fluctuations will
undergo rapid dispersion in the interface regions as they propagate
along the field.
When MHD waves pass through a strongly inhomogeneous background medium,
their properties can be altered in a number of ways.
Our understanding of these transformations has been shaped by concepts
from linear theory such as reflection \citep{heinemannolbert, chandranhollweg},
refraction \citep{stein, flaetal},
and mode coupling \citep{valley, poedts98, mecherimarsch}. 
Further suggestions for the transformation and dissipation of wave
energy---including phase mixing, shear-driven couplings, and various
instabilities---were summarized in Section 1.
It may also be the case that these kinds of mode conversions may be
responsible for producing low-frequency compressive waves in coronal
holes as observed recently by, e.g., \cite{krishnaprasad} and \cite{threlfall},
since such waves are unlikely to travel far into the
corona if they are produced only at the solar surface \citep{athaywhite}.

The spatial scale of the cross-field striations revealed by Comet Lovejoy
was roughly 4000 km at a heliocentric radius of $\sim$1.3 $R_{\odot}$.
This demonstrates the presence of magnetic flux tubes with diameters
at least an order of magnitude smaller than would exist if supergranular
``funnels'' were the smallest building blocks of coronal structure
(see, e.g., \cite{hackenberg00}).
The observed spatial scale of 4000 km falls within the range of
predictions of the perpendicular correlation length, or energy-containing
length, in recent models of MHD turbulence.
In these models, the turbulence is driven by the random-walk motions
of kiloGauss-field intergranular bright points on the solar surface.
Those bright points have horizontal scales of order 50--200 km in the
photosphere \citep{berger01, abramenko}.
In the open flux tube models of \cite{cranmervB05} and \cite{cranmer07},
the correlation length $\lambda_{\perp}$ expands
as the background field strength $B$ decreases, with
$\lambda_{\perp} \propto B^{-1/2}$.
At heights of 1.2--1.4 $R_{\odot}$, the \cite{cranmervB05}
model gave $\lambda_{\perp} \approx 4900$--6700 km, and the
\cite{cranmer07} model gave
$\lambda_{\perp} \approx 1250$--1700 km.
More recently, \cite{cranmervB12} updated this model to
include nonlinear effects that come to dominate in interplanetary space,
and at 1.2--1.4 $R_{\odot}$, the revised values of
$\lambda_{\perp}$ were found to be 2800--4300 km.
\cite{hollweg10} showed that this range of correlation lengths is
consistent with Faraday rotation fluctuations measured via radio sounding
of spacecraft signals through the inner corona.
The existence of density fluctuations on similar length scales is a
predicted feature in both ``passive scalar'' extensions of MHD turbulence
theory (e.g., \cite{harmoncoles, zank12})
and three-dimensional wave-driven models of the solar wind \citep{cranmer13}.

The images also allow us to test the theory of pickup ions.  We measure
a speed along the magnetic field of 154 \kms and a rate of spreading of
159 \kms in the plane of the sky.  After correction for projection effects,
the parallel speed is higher than expected from the comet velocity and
the magnetic field direction.  On the other hand, when those numbers
and the Alfv\'{e}n speed predicted by
the MAS MHD model of \cite{downs13} are put into the \cite{williamszank} 
pickup ion model, the rate of spreading along the field line is below
the predicted expansion rate.  The total kinetic energy of the oxygen ions
in the bispherical distribution is predicted to be about 60\% of the initial
oxygen atom kinetic energy, while 40\% goes into waves.  The observed
kinetic energy is about 2/3 the expected value, quite likely because
Coulomb equilibration with the protons from cometary $H_2O$ has
taken some energy from the oxygen.  That will not affect $V_{par}$,
but it will affect the rate of spreading along the magnetic field.

The MAS model predictions are in overall good agreement with the magnetic
field directions and densities inferred from the comet images, as reproted by
\cite{downs13}, provided that
shifts along the comet's path corresponding to a few degrees in solar 
longitude are permitted.  Considering that the models are based on a surface
magnetic field that is measured over the course of a Carrington rotation
and that the heating model currently used has substantial uncertainty, this
is quite good agreement. 

A detailed calculation of the electron temperature is beyond the scope of this
paper.  One complication is that the Alfv\'{e}n speed changes as cometary gas
is loaded onto the field lines, and another is that both cooling by excitation
and ionization of the O ions and Coulomb heating are likely to be important.
The Coulomb heating increases with charge state.  When the oxygen is singly ionized, 
the collision time is 200/$n_e$ seconds, and Coulomb collisions could not maintain 
the electrons at $10^6$ K against adiabatic expansion and radiative losses.  
However, the collision time drops as $Z^2$, so that Coulomb collisions could 
probably maintain that temperature in the region where the 171 \AA\/ emission 
arises.  In addition, some of the wave energy produced during isotropization could 
heat the electrons \citep{cairnszank}.

\acknowledgments

Patrick McCauley was supported by NSF SHINE Grant AGS-1259519
and NASA grant SPH02H1701R for SDO/AIA to the Smithsonian Astrophysical Observatory.
SDO is a NASA satellite and the AIA instrument team is led by Lockheed
Martin, with SAO as a major subcontractor.
Cooper Downs was supported by a subcontract from Lockheed Martin for NASA contract 
NNG04EA00, and by NASA Heliophysics Theory Program.  MHD simulations were conducted on
the NASA Pleiades and NSF Ranger supercomputers.  This work benefitted
from discussions at the workshop on the Science of Near-Sun Comets at
the International Space Sciences Institute.

{\it Facilities:} \facility{SDO (AIA)}.

\end{document}